\documentclass[letterpaper]{jpsj-suppl}
\usepackage{txfonts} 
\usepackage{url}

\title{Establishing mass spectrum of $S=-1$ hyperon resonances via a dynamical coupled-channels analysis of $K^-p$ reactions}

\author{Hiroyuki \textsc{Kamano}}

\inst{Research Center for Nuclear Physics, Osaka University, Ibaraki, Osaka 567-0047, Japan}

\email{kamano@rcnp.osaka-u.ac.jp}

\recdate{}

\abst{
We report our recent effort for the extraction of resonance parameters (complex pole mass and
residues etc.) associated with $\Lambda^*$ and $\Sigma^*$ hyperons.
This was accomplished via a comprehensive partial-wave analysis of
the data for $K^- p \to \bar K N, \pi \Sigma, \pi \Lambda, \eta \Lambda, K\Xi$ 
reactions from the thresholds up to $W=2.1$ GeV within a dynamical coupled-channels approach.
The results suggest a possible existence of new narrow $J^P=3/2^+$ $\Lambda$
resonance with pole mass $1671^{+2}_{-8} -i(5^{+11}_{-2})$ MeV, 
located close to the $\eta \Lambda$ threshold.
This resonance is found to be responsible for reproducing
the data for $K^-p \to \eta \Lambda$ differential cross sections near the threshold,
and thus the data seem favor its existence.
The extracted poles for $J^P=1/2^-$ $\Lambda$ resonances below the $\bar K N$ threshold,
including $\Lambda(1405)$, are also presented.
}

\kword{hyperons, coupled-channels equations, unitarity, resonances, partial-wave analysis, 
anti-kaon induced reactions, meson production reactions}

\begin{document}
\maketitle

\section{Introduction}

Over the past decades, $\Lambda$(1405) has attracted particular attention in the hadron 
physics because of its important role in understanding of the $\bar K N$ interactions 
at low energies.
However, of course $\Lambda$(1405) is not the only hyperon resonance with strangeness
$S=-1$ ($Y^*$), but actually a lot of $Y^*$ resonances 
are listed by PDG~\cite{pdg2014}.
In despite of its multitude of reported $Y^*$ resonances, most of them are much 
less understood than the nonstrangeness $N^*$ and $\Delta^*$ resonances. 
For example, even low-lying resonances are not well determined for $Y^*$.
In fact, most of the low-lying $\Sigma^*$ resonances with mass 
less than $\sim$ 1.7 GeV are poorly established, and even the spin-parity is not assigned 
for some of those resonances~(see page 1456 of Ref.~\cite{pdg2014}).
In addition, a possible existence of $J^P=1/2^-$ $\Lambda$ resonance 
with mass less than 1.4 GeV has been suggested (see, e.g., Ref.~\cite{ihjksy11}),
but its existence is still controversial.
Another example is that only the resonance parameters given by the
Breit-Wigner parametrization had been listed by PDG before 2012~\cite{pdg2012},
except for $\Sigma(1385)$ and $\Lambda(1520)$.
This is in contrast to the $N^*$ and $\Delta^*$ cases, where 
the resonances defined by poles of scattering amplitudes 
have also been extensively studied, and 
the resonance parameters extracted with both the pole definitions and 
the Breit-Wigner parametrizations are listed by PDG.

Recently, we have developed a dynamical coupled-channels (DCC) model for
$K^-p$ reactions \cite{kbp1,kbp2}, aiming at 
(a) extracting the $Y^*$ resonance parameters defined by poles of scattering amplitudes, 
(b) exploring the role of reaction dynamics in understanding of
the structure and dynamical origin etc. of $Y^*$ resonances, and 
(c) providing the elementary $\bar K N$ reaction amplitudes that 
can be used, e.g., for studying
the productions of hypernuclei and kaonic nuclei.
Our DCC model is based on multichannel scattering equations obeyed by 
the partial-wave amplitudes for $a \to b$ reactions:

\begin{equation}
T_{b,a} (p_b,p_a;W) = 
V_{b,a} (p_b,p_a;W)
+\sum_c \int q^2dq V_{b,c} (p_b,q;W) G_c(q;W) T_{c,a} (p_c,p_a;W),
\label{ls}
\end{equation}
with $G_c$ being the Green's function for the channel $c$, and $V_{b,a}$ being 
the transition potential given by
\begin{equation}
V_{b,a} (p_b,p_a;W) = 
v_{b,a} (p_b, p_a;W) + \sum_{Y^*_0} \frac{\Gamma_{b,Y^*_0} \Gamma_{a,Y^*_0}^\dag}{W-M_{Y^*_0}}.
\label{pot}
\end{equation}
 
Here, the first term of Eq.~(\ref{pot}), $v_{b,a}$, represents 
the hadron-exchange potential that contains only the ground-state mesons and baryons 
belonging to each flavor SU(3) multiplet, while the second term describes
the s-channel propagation of the ``bare'' excited hyperon states $Y^*_0$.
The channel space of the model is spanned
by the two-body $\bar KN$, $\pi \Sigma$, $\pi\Lambda$, $\eta \Lambda$, and $K\Xi$ states 
and also the three-body $\pi\pi\Lambda$ and $\pi\bar K N$ states that have 
the resonant $\pi\Sigma^*$ and $\bar K^* N$ components, respectively. 
The resulting coupled-channels equation~(\ref{ls}) satisfies the multichannel unitarity 
conditions and properly takes account of the dynamical effects arising from 
the off-shell rescattering processes.

The model parameters (bare masses and cut offs etc.) are determined by fitting 
the available data for the
$K^- p \to \bar K N, \pi \Sigma, \pi \Lambda, \eta \Lambda, K\Xi$ reactions 
from the thresholds up to $W=2.1$ GeV.
The data contains both unpolarized and polarized observables, and this results 
in fitting more than 17,000 data points.
From this analysis, we have determined the threshold parameters such as 
scattering lengths and effective ranges for the $MB$ scattering with 
$MB = \bar K N, \eta \Lambda, K\Xi$, and also the partial-wave amplitudes of
the $\bar K N \to \bar K N, \pi \Sigma, \pi \Lambda, \eta \Lambda, K\Xi$ reactions
not only for $S$ wave, but also $P$, $D$ and $F$ waves.
The details of our DCC analysis of
$K^- p \to \bar K N, \pi \Sigma, \pi \Lambda, \eta \Lambda, K\Xi$ reactions 
can be found in Refs.~\cite{kbp1}, and
in this contribution we present highlights of the $Y^*$ mass spectrum
extracted from our analysis~\cite{kbp2}.

\section{Highlights of Extracted $Y^*$ Spectrum via DCC Analysis of $K^- p$ Reactions}

\begin{figure}[t]
\centering
\includegraphics[clip,width=0.7\textwidth]{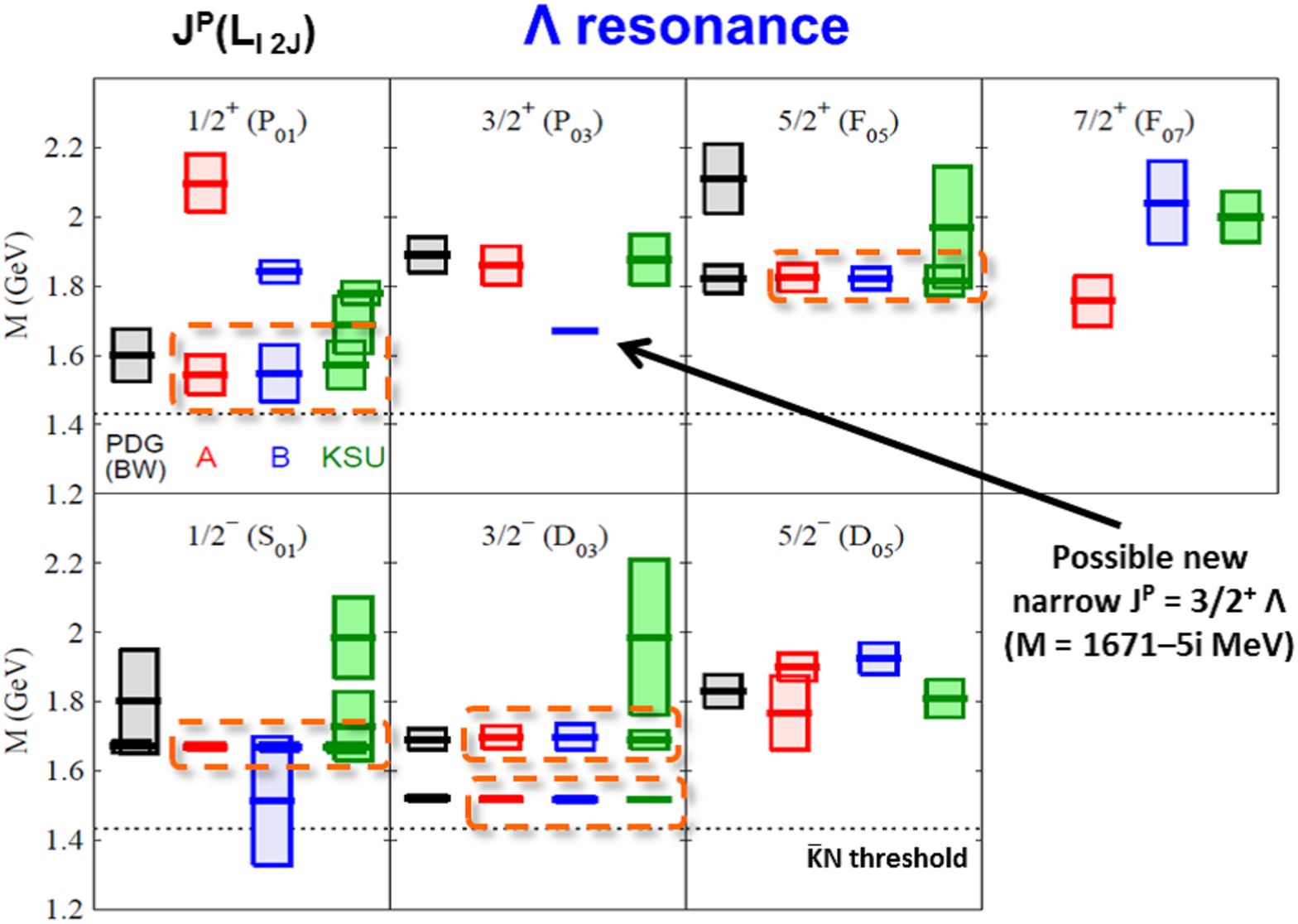}
\\
\vspace*{0.5em}
\includegraphics[clip,width=0.7\textwidth]{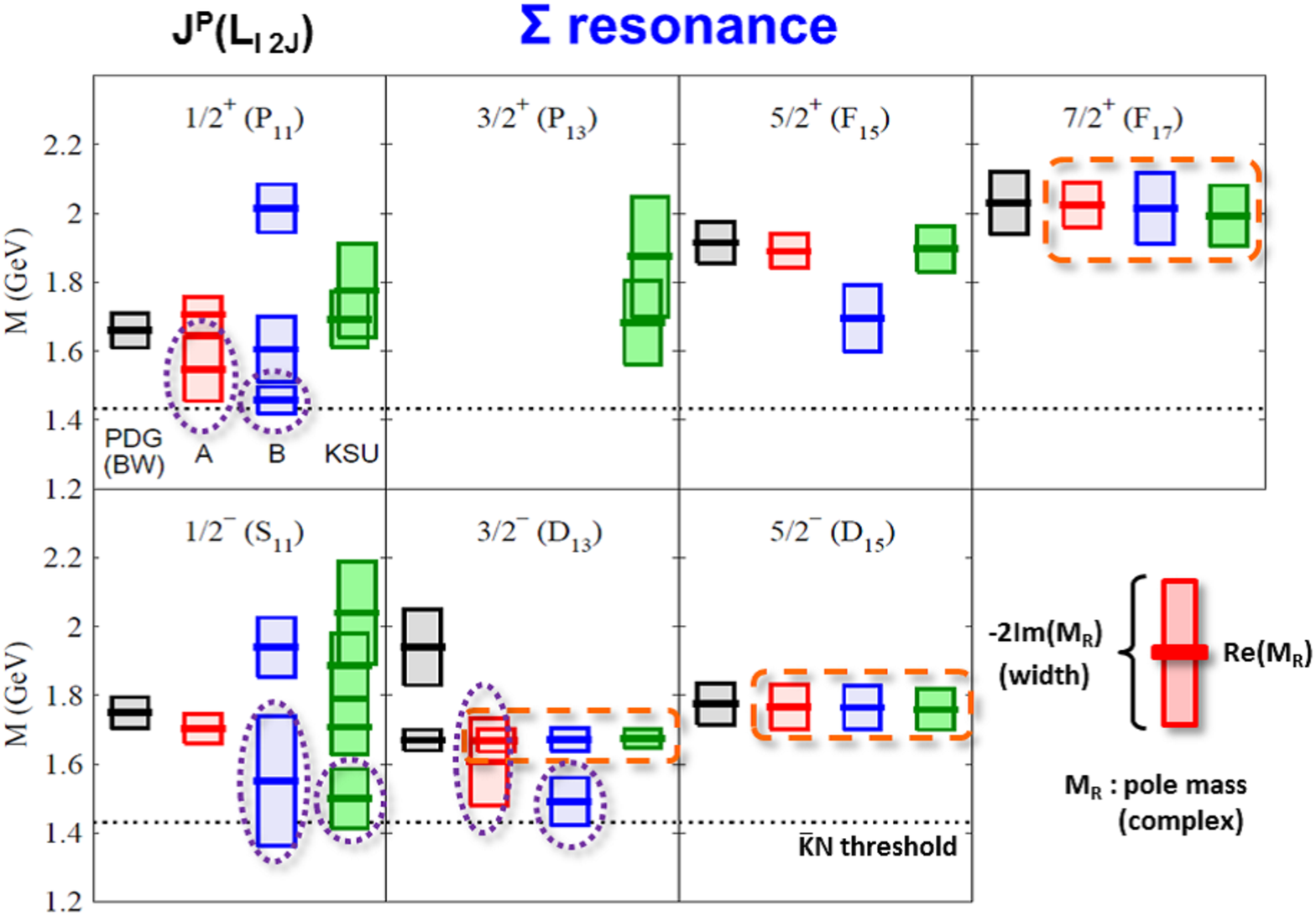}
\caption{
\label{spectrum}
Extracted mass spectra for $\Lambda$ (upper panels) and $\Sigma$ (lower panels) resonances.
Here only the resonances of which complex pole mass 
satisfies $m_{\bar K} + m_N \leq {\rm Re}(M_R) \leq 2.1$ GeV 
and $0 \leq -{\rm Im}(M_R) \leq 0.2$ GeV, are presented. 
The results are from Model A (red) and Model B (blue) of our DCC analysis~\cite{kbp1,kbp2}, 
and from the KSU analysis~\cite{zhang2013} (green).
The Breit-Wigner masses and widths of the four- and three-star resonances rated 
by PDG~\cite{pdg2014} (black) are also presented for a reference.
}
\end{figure}

Figure~\ref{spectrum} shows the comparison of $Y^*$ mass spectra extracted through 
comprehensive analyses of $K^- p$ reactions.
Here, the two spectra shown in red and blue are extracted from the two solutions
of our DCC analysis, named Model A and Model B, respectively~\cite{kbp1,kbp2}.
The appearance of the two solutions is because the available $K^-p$ reaction data 
are not sufficient to constrain our model parameters uniquely but it allowed us 
to have two distinct sets of our model parameters, yet both gives almost 
the same $\chi^2$ values.
The results from our two models and the one extracted by the KSU analysis~\cite{zhang2013}
agree very well for several resonances that are enclosed by orange dotted line
in Fig.~\ref{spectrum}.
However, in overall, the extracted results are still fluctuating
between our two models and the KSU analysis.
Again, this is because the existing $K^-p$ reaction data are not sufficient 
to constrain the analysis, 
and without new data this level of analysis dependence in the spectrum will not be avoidable.

It is interesting to see in Fig.~\ref{spectrum} that 
Model B has a very narrow $J^P=3/2^+$ $\Lambda$ resonance that is completely new.
The pole mass of this resonance is $M_R = 1671^{+2}_{-8}-i(5^{+11}_{-2})$ MeV, and 
it is located just above the $\eta \Lambda$ threshold.
Although at present this resonance is only seen in Model B, 
we have found in Refs.~\cite{kbp1,kbp2} that 
this new $J^P=3/2^+$ $\Lambda$ resonance is responsible for the reproduction of
the concave-up angular dependence of the $K^- p \to \eta \Lambda$ differential cross sections
near the threshold, and thus the existence of this resonance seems to be favored 
by the experimental data.
In fact, Model A does not reproduce the angular dependence of the 
$K^- p \to \eta \Lambda$ differential cross section data so well as compared to Model B, 
and those data were not included in the KSU analysis~\cite{zhang2013}.
For further confirmation, however,
one needs more extensive and accurate data for $K^- p \to \eta \Lambda $ reactions
near the threshold, including the recoil polarization $P$ and hopefully the spin-rotation 
angle $\beta$.

Another interesting thing is that a number of low-lying $\Sigma$ resonances located just above
the $\bar KN$ threshold are found 
(the resonances enclosed by purple dotted line in Fig.~\ref{spectrum}).
Their existence is still analysis dependent, but they may correspond to the one-star and 
two-star resonances listed by PDG~\cite{pdg2014}.
Again, much more data are definitely needed to establish these low-lying resonances.

\begin{figure}[t]
\centering
\includegraphics[clip,width=0.75\textwidth]{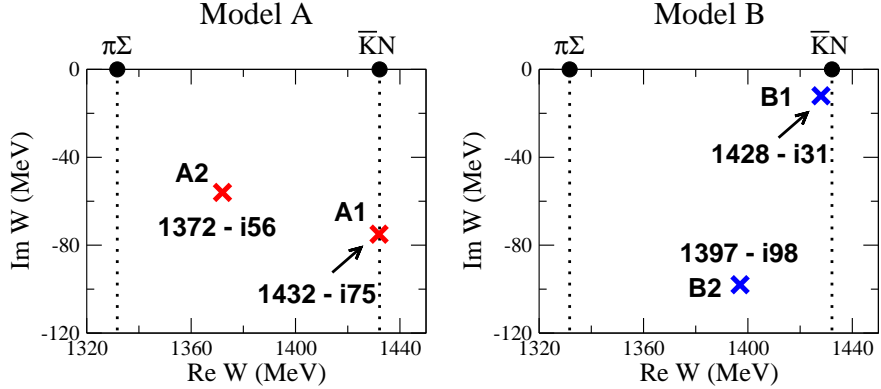}
\caption{
\label{s01}
``Predicted'' values of poles of $J^P= 1/2^-$ $\Lambda$ resonances located in the $\bar K N$
subthreshold region~\cite{kbp2}.
The result in the left (right) panel is from Model A (Model B)~\cite{kbp2}.
}
\end{figure}

Finally, we present the $J^P =1/2^-$ $\Lambda$ resonances lying below the $\bar K N$ threshold,
which are found from our two models (Fig.~\ref{s01}). 
Here we note that our current models were constructed by fitting only the $K^-p$ reactions, 
and thus the $\bar K N$ subthreshold region is out of the scope of our current analysis.
The extracted results should therefore be regarded as ``predictions'' from our current models
and are subject to change once our analysis is extended to the $\bar K N$ subthreshold region.
We see from Fig.~\ref{s01} that both models have two resonance poles, 
where the higher mass poles (A1 and B1) should correspond to $\Lambda(1405)$.
On the other hand, the existence of another resonance state (A2 and B2) with mass lower than
$\Lambda(1405)$ seems consistent with the one obtained in the so-called chiral unitary models
(see, e.g., Ref.~\cite{ihjksy11}).

\section{Summary and Future Works}

We have presented highlighted results for the $Y^*$ resonance mass spectrum
extracted via our DCC analysis of the data for the
$K^- p \to \bar K N, \pi \Sigma, \pi\Lambda, \eta \Lambda, K\Xi$ reactions 
over the wide energy range from the thresholds up to $W=2.1$ GeV.
The results suggest possible existences of the new narrow $J^P=3/2^+$ $\Lambda$ resonance
and unestablished low-lying $\Sigma^*$ resonances, and give ``predictions'' for 
$J^P=1/2^-$ $\Lambda$ resonances located below the $\bar K N$ threshold.
It should be emphasized that this kind of comprehensive analysis to extract $Y^*$ 
resonance parameters defined by poles of scattering amplitudes is for the first time 
within the dynamical-model approaches.
However, a visible analysis dependence is found to still exist for the extracted results, 
and this is because the current database of $K^- p$ reactions is far from complete.
We hope that the hadron beam facilities such as J-PARC measure these very fundamental 
reactions that are necessary for establishing $Y^*$ resonances.

To extend our analysis to the $\bar K N$ subthreshold region, we are now developing a model for
the $K^- d \to \pi Y N$ reactions, which are being measured by the J-PARC E31 
experiment~\cite{noumi}.
A strong point of our model is that it can properly take account of the higher partial-wave
contributions for the elementary processes, which is in contrast to most of the previous
studies where only the $S$-wave contribution is included.
The result of our extended analysis will be presented elsewhere.

Although it was not discussed in this contribution, our models can provide 
both on- and off-shell amplitudes for $MB \to M'B'$ reactions 
with $MB,M'B' = \bar K N, \pi \Sigma, \pi \Lambda, \eta \Lambda, K\Xi, 
\pi\Sigma^*, \bar K^* N$~\cite{kbp1}.
These elementary amplitudes can be an important input to 
the hypernuclei and kaonic-nuclei production reactions.
We hope to make applications of our approach to such nuclear-target reactions.
\\

The author thanks T.-S.~H.~Lee, S.~X.~Nakamura, and T.~Sato for their collaborations.
This work was supported by the JSPS KAKENHI Grant No. 25800149 and by
the HPCI Strategic Program (Field~5 ``The Origin of Matter and the Universe'') of MEXT of Japan.

\end{document}